\documentstyle[12pt]{article}

\title{Decay on several sorts of heterogeneous centers:
Special monodisperse approximation in the situation
of strong unsymmetry. 4. Numerical results for the
approximation of essential asymptotes}

\author{V.Kurasov}

\date{Victor.Kurasov@pobox.spbu.ru}

\begin{document}

\maketitle

This manuscript
directly continues \cite{Section1}, \cite{Section2}, \cite{Section3}.
All
definitions and formulas have to be taken from \cite{Section1}. The numerical
results for comparison with the total monodisperse approximation has
to be taken from \cite{Section2}.
The numerical results for the floating monodisperse approximation can
be taken from \cite{Section3}.

\section{Calculations}

We have to recall again   the system of the condensation equations. It can
be written in  the following form
$$
G = \int_0^z \exp(-G(x)) \theta_1(x) (z-x)^3 dx
$$
$$
\theta_1 = exp(-b \int_0^z \exp(-G(x)) dx)
$$
with a positive parameter $b$ and
have to estimate the error in
$$
N = \int_0^{\infty} \exp(-l G(x)) dx
$$
with some parameter $l$.

We shall solve this problem numerically  and compare our result with
the already formulated
models. In the model of the total monodisperse approximation we get
$$
N_A = \int_0^{\infty} \exp(-l G_A(x)) dx
$$
where
$G_A$ is
$$
G_A = \frac{1}{b} (1 - \exp(-b D)) x^3
$$
and the constant $D$ is given by
$$
D = \int_0^{\infty} \exp(-x^4 /4) dx = 1.28
$$
Numerical results are shown in \cite{Section2}.

In the
model of the floating monodisperse approximation we have to calculate
the integral
$$
N_B = \int_0^{\infty} \exp(-l G_B(x)) dx
$$
where
$G_B$ is
$$
G_B = \frac{1}{b} (1 - \exp(-b \int_0^{z/4} \exp(-x^4/4) dx )) z^3
$$
$$
G_B \approx \frac{1}{b} (1 - \exp(-b (\Theta(D-z/4) z/4 + \Theta(z/4-D) D) ))
z^3
$$
Numerical results are shown in \cite{Section3}.

It is very attractive to spread
the approximation for the last integral at small $z$ for all $z$ (as it
was done in the intermediate situation in \cite{Multidecay} when we solved
the algebraic equation on the parameters of the spectrum (in the intermediate
situation it is absolutely justified). Then we came to the  third
approximation
$$
N_C = \int_0^{\infty} \exp(-l G_C(x)) dx
$$
where
$G_C$ is
$$
G_C \approx \frac{1}{b} (1 - \exp(-b z/4 )) z^3
$$

This approximation will be called as "approximation of essential asymptotes".
The real advantage of this approximation is the absence of the exponential
nonlinearity. When this approximation will be introduced into equation
on the parameters of the condensation process there will be no numerical
difficulties to solve it.

We have tried all mentioned approximations for $b$ from $0.2$ up to $5.2$
with the step $0.2$ and for $l$ from $0.2$ up to $5.2$ with a step $0.2$.
We calculate the relative error in $N$. The results are drawn in fig.1
 for $N_C$ where the relative errors
are marked by  $r_3$.

We see that the relative errors of $N_B$ and $N_C$ are very small and
practically the same. One can not find the difference between fig.1 in
\cite{Section3} and
fig.1 here.

The maximum of errors in $N_B$ and $N_C$ lies near $l=0$. So, we have
to analyse the situation with small values of $l$. It was done in fig.2
 for $N_C$. We see that we can not
find the maximum error because it increases at small $b$. Then we
 have to calculate the situation with $b=0$.
Here we have to solve the following equation
$$
G = \int_0^{\infty} \exp(-G(x)) (z-x)^3 dx
$$
and to compare
$$
N = \int_0^{\infty} \exp(-l G) dx
$$
with
$$
N_A = \int_0^{\infty}
\exp(-l D z^3)        dz
$$
$$
N_B = \int_0^{\infty}
\exp(-l (\Theta(z/4 -D) D z^3 + \Theta(D-z/4) z^4 /4 )  )    dz
$$
$$
N_C = \int_0^{\infty}
\exp(-l  z^4 / 4) dz
$$
We can not put here $l=0$ directly.

The results are shown in fig.3. One can see one curve with two wings.
The upper wing corresponds to the error of $N_A$ and the lower corresponds
to the relative error in $N_B$ and $N_C$. At $l=0$ these wings
come together.
 We see that our hypothesis  (the worst situation for the floating
monodisperse approximation takes place when the first type heterogeneous
centers are unexhausted) is really true.

The worst situation is when $b$
is near zero and $l$ lies also near zero. Here we can use the total
monodisperse
approximation to estimate the error. It is clear that the relative error
in $N_A$ is greater than in $N_B$ (not in $N_C$). So, we can calculate
$r_1$, see that when $b$ goes to zero it decreases (it is clear also from
the physical reasons) and estimate the error $r_2$ at $b=0$, $l=0$ by
$R_1$ calculated at small $b$ and $l=0$. Then one can see that it is small.

An interesting problem is to see whether $N_B$ and $N_C$ are different
or no. Earlier we can not see the difference. In fig.4 one can see the
ratio $r_2/ r_3$ plotted at $b=0$ and can note that only for $l \approx
0.01 \div 0.02$ one can see the small difference. It means that to see
the difference
between these approximations the ratio between  the  scale  of  the
first type
centers nucleation and the scale of the second type centers nucleation
must be giant. Even at giant values the difference is small.

\pagebreak

\begin{picture}(350,350)
\put(0,0){\line(0,1){350}}
\put(0,0){\line(1,0){350}}
\put(350,0){\line(0,1){350}}
\put(0,350){\line(1,0){350}}
\put(152,206){.}
\put(158,208){.}
\put(164,210){.}
\put(170,212){.}
\put(176,214){.}
\put(182,215){.}
\put(188,217){.}
\put(194,218){.}
\put(200,219){.}
\put(206,219){.}
\put(212,220){.}
\put(218,220){.}
\put(224,221){.}
\put(230,221){.}
\put(236,221){.}
\put(242,221){.}
\put(248,220){.}
\put(254,220){.}
\put(260,220){.}
\put(266,219){.}
\put(272,219){.}
\put(278,218){.}
\put(284,218){.}
\put(290,217){.}
\put(296,216){.}
\put(302,216){.}
\put(148,199){.}
\put(154,201){.}
\put(160,203){.}
\put(166,205){.}
\put(172,207){.}
\put(178,208){.}
\put(184,210){.}
\put(190,211){.}
\put(196,212){.}
\put(202,213){.}
\put(208,214){.}
\put(214,215){.}
\put(220,215){.}
\put(226,216){.}
\put(232,216){.}
\put(238,216){.}
\put(244,216){.}
\put(250,216){.}
\put(256,216){.}
\put(262,216){.}
\put(268,216){.}
\put(274,215){.}
\put(280,215){.}
\put(286,214){.}
\put(292,214){.}
\put(298,214){.}
\put(143,194){.}
\put(149,196){.}
\put(155,198){.}
\put(161,199){.}
\put(167,201){.}
\put(173,203){.}
\put(179,204){.}
\put(185,205){.}
\put(191,207){.}
\put(197,208){.}
\put(203,208){.}
\put(209,209){.}
\put(215,210){.}
\put(221,210){.}
\put(227,211){.}
\put(233,211){.}
\put(239,211){.}
\put(245,212){.}
\put(251,212){.}
\put(257,212){.}
\put(263,211){.}
\put(269,211){.}
\put(275,211){.}
\put(281,211){.}
\put(287,211){.}
\put(293,210){.}
\put(139,189){.}
\put(145,191){.}
\put(151,193){.}
\put(157,194){.}
\put(163,196){.}
\put(169,198){.}
\put(175,199){.}
\put(181,200){.}
\put(187,201){.}
\put(193,202){.}
\put(199,203){.}
\put(205,204){.}
\put(211,205){.}
\put(217,206){.}
\put(223,206){.}
\put(229,206){.}
\put(235,207){.}
\put(241,207){.}
\put(247,207){.}
\put(253,207){.}
\put(259,207){.}
\put(265,207){.}
\put(271,207){.}
\put(277,207){.}
\put(283,207){.}
\put(289,206){.}
\put(135,184){.}
\put(141,186){.}
\put(147,188){.}
\put(153,190){.}
\put(159,191){.}
\put(165,193){.}
\put(171,194){.}
\put(177,195){.}
\put(183,197){.}
\put(189,198){.}
\put(195,199){.}
\put(201,199){.}
\put(207,200){.}
\put(213,201){.}
\put(219,201){.}
\put(225,202){.}
\put(231,202){.}
\put(237,202){.}
\put(243,203){.}
\put(249,203){.}
\put(255,203){.}
\put(261,203){.}
\put(267,203){.}
\put(273,203){.}
\put(279,203){.}
\put(285,202){.}
\put(131,180){.}
\put(137,182){.}
\put(143,183){.}
\put(149,185){.}
\put(155,187){.}
\put(161,188){.}
\put(167,189){.}
\put(173,191){.}
\put(179,192){.}
\put(185,193){.}
\put(191,194){.}
\put(197,195){.}
\put(203,195){.}
\put(209,196){.}
\put(215,197){.}
\put(221,197){.}
\put(227,198){.}
\put(233,198){.}
\put(239,198){.}
\put(245,198){.}
\put(251,199){.}
\put(257,199){.}
\put(263,199){.}
\put(269,199){.}
\put(275,198){.}
\put(281,198){.}
\put(126,175){.}
\put(132,177){.}
\put(138,179){.}
\put(144,180){.}
\put(150,182){.}
\put(156,183){.}
\put(162,185){.}
\put(168,186){.}
\put(174,187){.}
\put(180,188){.}
\put(186,189){.}
\put(192,190){.}
\put(198,191){.}
\put(204,191){.}
\put(210,192){.}
\put(216,193){.}
\put(222,193){.}
\put(228,193){.}
\put(234,194){.}
\put(240,194){.}
\put(246,194){.}
\put(252,194){.}
\put(258,194){.}
\put(264,194){.}
\put(270,194){.}
\put(276,194){.}
\put(122,171){.}
\put(128,173){.}
\put(134,174){.}
\put(140,176){.}
\put(146,177){.}
\put(152,179){.}
\put(158,180){.}
\put(164,181){.}
\put(170,182){.}
\put(176,184){.}
\put(182,184){.}
\put(188,185){.}
\put(194,186){.}
\put(200,187){.}
\put(206,187){.}
\put(212,188){.}
\put(218,189){.}
\put(224,189){.}
\put(230,189){.}
\put(236,190){.}
\put(242,190){.}
\put(248,190){.}
\put(254,190){.}
\put(260,190){.}
\put(266,190){.}
\put(272,190){.}
\put(118,167){.}
\put(124,168){.}
\put(130,170){.}
\put(136,171){.}
\put(142,173){.}
\put(148,174){.}
\put(154,176){.}
\put(160,177){.}
\put(166,178){.}
\put(172,179){.}
\put(178,180){.}
\put(184,181){.}
\put(190,182){.}
\put(196,182){.}
\put(202,183){.}
\put(208,184){.}
\put(214,184){.}
\put(220,184){.}
\put(226,185){.}
\put(232,185){.}
\put(238,185){.}
\put(244,185){.}
\put(250,186){.}
\put(256,186){.}
\put(262,186){.}
\put(268,186){.}
\put(114,162){.}
\put(120,164){.}
\put(126,166){.}
\put(132,167){.}
\put(138,168){.}
\put(144,170){.}
\put(150,171){.}
\put(156,172){.}
\put(162,173){.}
\put(168,174){.}
\put(174,175){.}
\put(180,176){.}
\put(186,177){.}
\put(192,178){.}
\put(198,178){.}
\put(204,179){.}
\put(210,180){.}
\put(216,180){.}
\put(222,180){.}
\put(228,181){.}
\put(234,181){.}
\put(240,181){.}
\put(246,181){.}
\put(252,181){.}
\put(258,181){.}
\put(264,181){.}
\put(109,158){.}
\put(115,160){.}
\put(121,161){.}
\put(127,163){.}
\put(133,164){.}
\put(139,165){.}
\put(145,167){.}
\put(151,168){.}
\put(157,169){.}
\put(163,170){.}
\put(169,171){.}
\put(175,172){.}
\put(181,173){.}
\put(187,173){.}
\put(193,174){.}
\put(199,175){.}
\put(205,175){.}
\put(211,176){.}
\put(217,176){.}
\put(223,176){.}
\put(229,177){.}
\put(235,177){.}
\put(241,177){.}
\put(247,177){.}
\put(253,177){.}
\put(259,177){.}
\put(105,154){.}
\put(111,155){.}
\put(117,157){.}
\put(123,158){.}
\put(129,160){.}
\put(135,161){.}
\put(141,162){.}
\put(147,163){.}
\put(153,164){.}
\put(159,165){.}
\put(165,166){.}
\put(171,167){.}
\put(177,168){.}
\put(183,169){.}
\put(189,170){.}
\put(195,170){.}
\put(201,171){.}
\put(207,171){.}
\put(213,172){.}
\put(219,172){.}
\put(225,172){.}
\put(231,172){.}
\put(237,173){.}
\put(243,173){.}
\put(249,173){.}
\put(255,173){.}
\put(101,149){.}
\put(107,151){.}
\put(113,152){.}
\put(119,154){.}
\put(125,155){.}
\put(131,157){.}
\put(137,158){.}
\put(143,159){.}
\put(149,160){.}
\put(155,161){.}
\put(161,162){.}
\put(167,163){.}
\put(173,164){.}
\put(179,164){.}
\put(185,165){.}
\put(191,166){.}
\put(197,166){.}
\put(203,167){.}
\put(209,167){.}
\put(215,167){.}
\put(221,168){.}
\put(227,168){.}
\put(233,168){.}
\put(239,168){.}
\put(245,168){.}
\put(251,169){.}
\put(97,145){.}
\put(103,147){.}
\put(109,148){.}
\put(115,150){.}
\put(121,151){.}
\put(127,152){.}
\put(133,153){.}
\put(139,155){.}
\put(145,156){.}
\put(151,157){.}
\put(157,158){.}
\put(163,158){.}
\put(169,159){.}
\put(175,160){.}
\put(181,161){.}
\put(187,161){.}
\put(193,162){.}
\put(199,162){.}
\put(205,163){.}
\put(211,163){.}
\put(217,163){.}
\put(223,164){.}
\put(229,164){.}
\put(235,164){.}
\put(241,164){.}
\put(247,164){.}
\put(92,141){.}
\put(98,142){.}
\put(104,144){.}
\put(110,145){.}
\put(116,147){.}
\put(122,148){.}
\put(128,149){.}
\put(134,150){.}
\put(140,151){.}
\put(146,152){.}
\put(152,153){.}
\put(158,154){.}
\put(164,155){.}
\put(170,156){.}
\put(176,156){.}
\put(182,157){.}
\put(188,157){.}
\put(194,158){.}
\put(200,158){.}
\put(206,159){.}
\put(212,159){.}
\put(218,159){.}
\put(224,160){.}
\put(230,160){.}
\put(236,160){.}
\put(242,160){.}
\put(88,137){.}
\put(94,138){.}
\put(100,139){.}
\put(106,141){.}
\put(112,142){.}
\put(118,143){.}
\put(124,145){.}
\put(130,146){.}
\put(136,147){.}
\put(142,148){.}
\put(148,149){.}
\put(154,150){.}
\put(160,150){.}
\put(166,151){.}
\put(172,152){.}
\put(178,152){.}
\put(184,153){.}
\put(190,154){.}
\put(196,154){.}
\put(202,154){.}
\put(208,155){.}
\put(214,155){.}
\put(220,155){.}
\put(226,155){.}
\put(232,156){.}
\put(238,156){.}
\put(84,132){.}
\put(90,134){.}
\put(96,135){.}
\put(102,137){.}
\put(108,138){.}
\put(114,139){.}
\put(120,140){.}
\put(126,141){.}
\put(132,142){.}
\put(138,143){.}
\put(144,144){.}
\put(150,145){.}
\put(156,146){.}
\put(162,147){.}
\put(168,147){.}
\put(174,148){.}
\put(180,149){.}
\put(186,149){.}
\put(192,150){.}
\put(198,150){.}
\put(204,150){.}
\put(210,151){.}
\put(216,151){.}
\put(222,151){.}
\put(228,151){.}
\put(234,151){.}
\put(80,128){.}
\put(86,129){.}
\put(92,131){.}
\put(98,132){.}
\put(104,134){.}
\put(110,135){.}
\put(116,136){.}
\put(122,137){.}
\put(128,138){.}
\put(134,139){.}
\put(140,140){.}
\put(146,141){.}
\put(152,142){.}
\put(158,142){.}
\put(164,143){.}
\put(170,144){.}
\put(176,144){.}
\put(182,145){.}
\put(188,145){.}
\put(194,146){.}
\put(200,146){.}
\put(206,146){.}
\put(212,147){.}
\put(218,147){.}
\put(224,147){.}
\put(230,147){.}
\put(75,124){.}
\put(81,125){.}
\put(87,127){.}
\put(93,128){.}
\put(99,129){.}
\put(105,130){.}
\put(111,132){.}
\put(117,133){.}
\put(123,134){.}
\put(129,135){.}
\put(135,136){.}
\put(141,136){.}
\put(147,137){.}
\put(153,138){.}
\put(159,139){.}
\put(165,139){.}
\put(171,140){.}
\put(177,140){.}
\put(183,141){.}
\put(189,141){.}
\put(195,142){.}
\put(201,142){.}
\put(207,142){.}
\put(213,142){.}
\put(219,143){.}
\put(225,143){.}
\put(71,120){.}
\put(77,121){.}
\put(83,122){.}
\put(89,124){.}
\put(95,125){.}
\put(101,126){.}
\put(107,127){.}
\put(113,128){.}
\put(119,129){.}
\put(125,130){.}
\put(131,131){.}
\put(137,132){.}
\put(143,133){.}
\put(149,134){.}
\put(155,134){.}
\put(161,135){.}
\put(167,136){.}
\put(173,136){.}
\put(179,136){.}
\put(185,137){.}
\put(191,137){.}
\put(197,138){.}
\put(203,138){.}
\put(209,138){.}
\put(215,138){.}
\put(221,138){.}
\put(67,115){.}
\put(73,117){.}
\put(79,118){.}
\put(85,119){.}
\put(91,121){.}
\put(97,122){.}
\put(103,123){.}
\put(109,124){.}
\put(115,125){.}
\put(121,126){.}
\put(127,127){.}
\put(133,128){.}
\put(139,129){.}
\put(145,129){.}
\put(151,130){.}
\put(157,131){.}
\put(163,131){.}
\put(169,132){.}
\put(175,132){.}
\put(181,133){.}
\put(187,133){.}
\put(193,133){.}
\put(199,134){.}
\put(205,134){.}
\put(211,134){.}
\put(217,134){.}
\put(63,111){.}
\put(69,112){.}
\put(75,114){.}
\put(81,115){.}
\put(87,116){.}
\put(93,117){.}
\put(99,119){.}
\put(105,120){.}
\put(111,121){.}
\put(117,122){.}
\put(123,123){.}
\put(129,123){.}
\put(135,124){.}
\put(141,125){.}
\put(147,126){.}
\put(153,126){.}
\put(159,127){.}
\put(165,127){.}
\put(171,128){.}
\put(177,128){.}
\put(183,129){.}
\put(189,129){.}
\put(195,129){.}
\put(201,129){.}
\put(207,130){.}
\put(213,130){.}
\put(58,107){.}
\put(64,108){.}
\put(70,109){.}
\put(76,111){.}
\put(82,112){.}
\put(88,113){.}
\put(94,114){.}
\put(100,115){.}
\put(106,116){.}
\put(112,117){.}
\put(118,118){.}
\put(124,119){.}
\put(130,120){.}
\put(136,121){.}
\put(142,121){.}
\put(148,122){.}
\put(154,122){.}
\put(160,123){.}
\put(166,123){.}
\put(172,124){.}
\put(178,124){.}
\put(184,125){.}
\put(190,125){.}
\put(196,125){.}
\put(202,125){.}
\put(208,126){.}
\put(54,103){.}
\put(60,104){.}
\put(66,105){.}
\put(72,106){.}
\put(78,108){.}
\put(84,109){.}
\put(90,110){.}
\put(96,111){.}
\put(102,112){.}
\put(108,113){.}
\put(114,114){.}
\put(120,115){.}
\put(126,115){.}
\put(132,116){.}
\put(138,117){.}
\put(144,118){.}
\put(150,118){.}
\put(156,119){.}
\put(162,119){.}
\put(168,120){.}
\put(174,120){.}
\put(180,120){.}
\put(186,121){.}
\put(192,121){.}
\put(198,121){.}
\put(204,121){.}
\put(50,98){.}
\put(56,100){.}
\put(62,101){.}
\put(68,102){.}
\put(74,103){.}
\put(80,105){.}
\put(86,106){.}
\put(92,107){.}
\put(98,108){.}
\put(104,109){.}
\put(110,109){.}
\put(116,110){.}
\put(122,111){.}
\put(128,112){.}
\put(134,113){.}
\put(140,113){.}
\put(146,114){.}
\put(152,114){.}
\put(158,115){.}
\put(164,115){.}
\put(170,116){.}
\put(176,116){.}
\put(182,116){.}
\put(188,117){.}
\put(194,117){.}
\put(200,117){.}
\put(46,94){.}
\put(52,95){.}
\put(58,97){.}
\put(64,98){.}
\put(70,99){.}
\put(76,100){.}
\put(82,101){.}
\put(88,102){.}
\put(94,103){.}
\put(100,104){.}
\put(106,105){.}
\put(112,106){.}
\put(118,107){.}
\put(124,108){.}
\put(130,108){.}
\put(136,109){.}
\put(142,109){.}
\put(148,110){.}
\put(154,110){.}
\put(160,111){.}
\put(166,111){.}
\put(172,112){.}
\put(178,112){.}
\put(184,112){.}
\put(190,112){.}
\put(196,113){.}
\put(150,200){\vector(0,1){100}}
\put(150,200){\vector(1,0){170}}
\put(150,200){\vector(-1,-1){120}}
\put(140,300){$r_3$}
\put(335,205){$b$}
\put(20,70){$l$}
\put(40,90){\line(1,0){156}}
\put(40,90){\line(0,1){4}}
\put(20,90){5.2}
\put(306,200){\line(-1,-1){110}}
\put(306,200){\line(0,1){20}}
\put(306,215){5.2}
\put(196,90){\line(0,1){23}}
\put(125,245){0.1}
\end{picture}

{ \it

\begin{center}
Fig.1
\end{center}

 The relative error of $N_C$ drawn as the function of $l$ and $b$.
Parameter $l$ goes from $0.2$ up to $5.2$ with a step $0.2$.
Parameter $b$ goes from $0.2$ up to $5.2$ with a step $0.2$.

One can see the maximum at small $l$ and moderate $b$. One can not separate
$N_B$ and $N_C$ according to fig.1 in \cite{Section3} and fig.1.

}

\pagebreak

\begin{picture}(350,350)
\put(0,0){\line(0,1){350}}
\put(0,0){\line(1,0){350}}
\put(350,0){\line(0,1){350}}
\put(0,350){\line(1,0){350}}
\put(145,224){.}
\put(151,224){.}
\put(157,222){.}
\put(163,221){.}
\put(169,220){.}
\put(175,218){.}
\put(181,217){.}
\put(187,216){.}
\put(193,214){.}
\put(199,213){.}
\put(205,211){.}
\put(211,210){.}
\put(217,209){.}
\put(223,208){.}
\put(229,206){.}
\put(235,205){.}
\put(241,204){.}
\put(247,203){.}
\put(253,202){.}
\put(259,202){.}
\put(265,201){.}
\put(271,200){.}
\put(277,199){.}
\put(283,198){.}
\put(289,198){.}
\put(295,197){.}
\put(135,208){.}
\put(141,209){.}
\put(147,209){.}
\put(153,209){.}
\put(159,209){.}
\put(165,208){.}
\put(171,208){.}
\put(177,207){.}
\put(183,206){.}
\put(189,205){.}
\put(195,204){.}
\put(201,203){.}
\put(207,201){.}
\put(213,200){.}
\put(219,199){.}
\put(225,198){.}
\put(231,197){.}
\put(237,196){.}
\put(243,195){.}
\put(249,194){.}
\put(255,193){.}
\put(261,192){.}
\put(267,191){.}
\put(273,191){.}
\put(279,190){.}
\put(285,189){.}
\put(124,194){.}
\put(130,195){.}
\put(136,196){.}
\put(142,196){.}
\put(148,197){.}
\put(154,197){.}
\put(160,197){.}
\put(166,196){.}
\put(172,196){.}
\put(178,195){.}
\put(184,194){.}
\put(190,193){.}
\put(196,192){.}
\put(202,191){.}
\put(208,190){.}
\put(214,189){.}
\put(220,188){.}
\put(226,187){.}
\put(232,186){.}
\put(238,185){.}
\put(244,184){.}
\put(250,183){.}
\put(256,183){.}
\put(262,182){.}
\put(268,181){.}
\put(274,180){.}
\put(114,180){.}
\put(120,182){.}
\put(126,183){.}
\put(132,184){.}
\put(138,185){.}
\put(144,185){.}
\put(150,185){.}
\put(156,185){.}
\put(162,185){.}
\put(168,185){.}
\put(174,184){.}
\put(180,183){.}
\put(186,182){.}
\put(192,182){.}
\put(198,181){.}
\put(204,180){.}
\put(210,179){.}
\put(216,178){.}
\put(222,177){.}
\put(228,176){.}
\put(234,175){.}
\put(240,174){.}
\put(246,173){.}
\put(252,172){.}
\put(258,172){.}
\put(264,171){.}
\put(103,167){.}
\put(109,169){.}
\put(115,171){.}
\put(121,172){.}
\put(127,173){.}
\put(133,173){.}
\put(139,174){.}
\put(145,174){.}
\put(151,174){.}
\put(157,174){.}
\put(163,173){.}
\put(169,173){.}
\put(175,172){.}
\put(181,171){.}
\put(187,171){.}
\put(193,170){.}
\put(199,169){.}
\put(205,168){.}
\put(211,167){.}
\put(217,166){.}
\put(223,165){.}
\put(229,164){.}
\put(235,164){.}
\put(241,163){.}
\put(247,162){.}
\put(253,161){.}
\put(92,155){.}
\put(98,157){.}
\put(104,158){.}
\put(110,160){.}
\put(116,161){.}
\put(122,162){.}
\put(128,162){.}
\put(134,163){.}
\put(140,163){.}
\put(146,163){.}
\put(152,163){.}
\put(158,162){.}
\put(164,162){.}
\put(170,161){.}
\put(176,161){.}
\put(182,160){.}
\put(188,159){.}
\put(194,158){.}
\put(200,157){.}
\put(206,156){.}
\put(212,155){.}
\put(218,155){.}
\put(224,154){.}
\put(230,153){.}
\put(236,152){.}
\put(242,151){.}
\put(82,143){.}
\put(88,145){.}
\put(94,147){.}
\put(100,148){.}
\put(106,149){.}
\put(112,150){.}
\put(118,151){.}
\put(124,152){.}
\put(130,152){.}
\put(136,152){.}
\put(142,152){.}
\put(148,152){.}
\put(154,151){.}
\put(160,151){.}
\put(166,150){.}
\put(172,150){.}
\put(178,149){.}
\put(184,148){.}
\put(190,147){.}
\put(196,146){.}
\put(202,145){.}
\put(208,145){.}
\put(214,144){.}
\put(220,143){.}
\put(226,142){.}
\put(232,141){.}
\put(71,131){.}
\put(77,133){.}
\put(83,135){.}
\put(89,137){.}
\put(95,138){.}
\put(101,139){.}
\put(107,140){.}
\put(113,141){.}
\put(119,141){.}
\put(125,141){.}
\put(131,141){.}
\put(137,141){.}
\put(143,141){.}
\put(149,140){.}
\put(155,140){.}
\put(161,139){.}
\put(167,138){.}
\put(173,138){.}
\put(179,137){.}
\put(185,136){.}
\put(191,135){.}
\put(197,135){.}
\put(203,134){.}
\put(209,133){.}
\put(215,132){.}
\put(221,131){.}
\put(61,120){.}
\put(67,122){.}
\put(73,124){.}
\put(79,125){.}
\put(85,127){.}
\put(91,128){.}
\put(97,129){.}
\put(103,130){.}
\put(109,130){.}
\put(115,130){.}
\put(121,130){.}
\put(127,130){.}
\put(133,130){.}
\put(139,130){.}
\put(145,129){.}
\put(151,129){.}
\put(157,128){.}
\put(163,127){.}
\put(169,127){.}
\put(175,126){.}
\put(181,125){.}
\put(187,124){.}
\put(193,124){.}
\put(199,123){.}
\put(205,122){.}
\put(211,121){.}
\put(50,108){.}
\put(56,111){.}
\put(62,112){.}
\put(68,114){.}
\put(74,116){.}
\put(80,117){.}
\put(86,118){.}
\put(92,119){.}
\put(98,119){.}
\put(104,119){.}
\put(110,120){.}
\put(116,120){.}
\put(122,119){.}
\put(128,119){.}
\put(134,119){.}
\put(140,118){.}
\put(146,118){.}
\put(152,117){.}
\put(158,116){.}
\put(164,116){.}
\put(170,115){.}
\put(176,114){.}
\put(182,113){.}
\put(188,113){.}
\put(194,112){.}
\put(200,111){.}
\put(39,97){.}
\put(45,99){.}
\put(51,101){.}
\put(57,103){.}
\put(63,104){.}
\put(69,106){.}
\put(75,107){.}
\put(81,108){.}
\put(87,108){.}
\put(93,109){.}
\put(99,109){.}
\put(105,109){.}
\put(111,109){.}
\put(117,109){.}
\put(123,108){.}
\put(129,108){.}
\put(135,107){.}
\put(141,107){.}
\put(147,106){.}
\put(153,105){.}
\put(159,105){.}
\put(165,104){.}
\put(171,103){.}
\put(177,102){.}
\put(183,102){.}
\put(189,101){.}
\put(150,200){\vector(0,1){100}}
\put(150,200){\vector(1,0){170}}
\put(150,200){\vector(-1,-1){120}}
\put(140,300){$r_3$}
\put(335,205){$b$}
\put(20,70){$l$}
\put(33,83){\line(1,0){156}}
\put(33,83){\line(0,1){14}}
\put(13,83){0.1}
\put(306,200){\line(-1,-1){117}}
\put(306,200){\line(0,1){8}}
\put(311,215){5.2}
\put(189,83){\line(0,1){18}}
\put(125,245){0.1}
\end{picture}

\begin{center}
Fig.2
\end{center}

{ \it
 The relative error of $N_C$ drawn as the function of $l$ and $b$.
Parameter $l$ goes from $0.01$ up to $0.11$ with a step $0.01$.
Parameter $b$ goes from $0.2$ up to $5.2$ with a step $0.2$.

One can see the maximum at small $l$ and small $b$. One can note that
now the values of $b$ corresponding to maximum of the relative errors
become small. One can not separate
$N_B$ and $N_C$ according to fig.2 in \cite{Section3} and fig.2 here.

}

\pagebreak

\begin{picture}(350,350)
\put(0,0){\line(0,1){350}}
\put(0,0){\line(1,0){350}}
\put(350,0){\line(0,1){350}}
\put(0,350){\line(1,0){350}}
\put(26,109){.}
\put(26,113){.}
\put(26,104){.}
\put(26,126){.}
\put(26,90){.}
\put(26,89){.}
\put(27,136){.}
\put(27,78){.}
\put(27,78){.}
\put(27,144){.}
\put(27,71){.}
\put(27,71){.}
\put(28,151){.}
\put(28,65){.}
\put(28,65){.}
\put(29,156){.}
\put(29,61){.}
\put(29,61){.}
\put(29,160){.}
\put(29,58){.}
\put(29,58){.}
\put(30,164){.}
\put(30,56){.}
\put(30,56){.}
\put(30,168){.}
\put(30,54){.}
\put(30,54){.}
\put(31,171){.}
\put(31,52){.}
\put(31,52){.}
\put(32,177){.}
\put(32,49){.}
\put(32,49){.}
\put(33,181){.}
\put(33,47){.}
\put(33,47){.}
\put(35,186){.}
\put(35,45){.}
\put(35,45){.}
\put(36,189){.}
\put(36,44){.}
\put(36,44){.}
\put(37,193){.}
\put(37,43){.}
\put(37,43){.}
\put(38,196){.}
\put(38,42){.}
\put(38,42){.}
\put(39,198){.}
\put(39,41){.}
\put(39,41){.}
\put(41,201){.}
\put(41,40){.}
\put(41,40){.}
\put(42,203){.}
\put(42,39){.}
\put(42,39){.}
\put(43,205){.}
\put(43,39){.}
\put(43,39){.}
\put(45,208){.}
\put(45,38){.}
\put(45,38){.}
\put(47,211){.}
\put(47,37){.}
\put(47,37){.}
\put(48,214){.}
\put(48,37){.}
\put(48,37){.}
\put(50,216){.}
\put(50,36){.}
\put(50,36){.}
\put(52,218){.}
\put(52,36){.}
\put(52,36){.}
\put(54,220){.}
\put(54,36){.}
\put(54,36){.}
\put(56,222){.}
\put(56,35){.}
\put(56,35){.}
\put(58,225){.}
\put(58,35){.}
\put(58,35){.}
\put(60,227){.}
\put(60,35){.}
\put(60,35){.}
\put(63,229){.}
\put(63,34){.}
\put(63,34){.}
\put(65,231){.}
\put(65,34){.}
\put(65,34){.}
\put(68,233){.}
\put(68,34){.}
\put(68,34){.}
\put(71,235){.}
\put(71,34){.}
\put(71,34){.}
\put(74,237){.}
\put(74,33){.}
\put(74,33){.}
\put(77,239){.}
\put(77,33){.}
\put(77,33){.}
\put(80,241){.}
\put(80,33){.}
\put(80,33){.}
\put(83,243){.}
\put(83,33){.}
\put(83,33){.}
\put(87,244){.}
\put(87,33){.}
\put(87,33){.}
\put(90,246){.}
\put(90,33){.}
\put(90,33){.}
\put(94,248){.}
\put(94,32){.}
\put(94,32){.}
\put(98,250){.}
\put(98,32){.}
\put(98,32){.}
\put(102,251){.}
\put(102,32){.}
\put(102,32){.}
\put(107,253){.}
\put(107,32){.}
\put(107,32){.}
\put(111,255){.}
\put(111,32){.}
\put(111,32){.}
\put(116,256){.}
\put(116,32){.}
\put(116,32){.}
\put(122,258){.}
\put(122,32){.}
\put(122,32){.}
\put(127,260){.}
\put(127,32){.}
\put(127,32){.}
\put(133,262){.}
\put(133,32){.}
\put(133,32){.}
\put(139,263){.}
\put(139,32){.}
\put(139,32){.}
\put(146,265){.}
\put(146,32){.}
\put(146,32){.}
\put(152,267){.}
\put(152,32){.}
\put(152,32){.}
\put(159,268){.}
\put(159,32){.}
\put(159,32){.}
\put(167,270){.}
\put(167,32){.}
\put(167,32){.}
\put(174,271){.}
\put(174,32){.}
\put(174,32){.}
\put(182,273){.}
\put(182,32){.}
\put(182,32){.}
\put(191,275){.}
\put(191,31){.}
\put(191,31){.}
\put(200,276){.}
\put(200,31){.}
\put(200,31){.}
\put(209,278){.}
\put(209,31){.}
\put(209,31){.}
\put(219,279){.}
\put(219,31){.}
\put(219,31){.}
\put(229,281){.}
\put(229,31){.}
\put(229,31){.}
\put(240,282){.}
\put(240,32){.}
\put(240,32){.}
\put(251,284){.}
\put(251,32){.}
\put(251,32){.}
\put(263,285){.}
\put(263,32){.}
\put(263,32){.}
\put(276,287){.}
\put(276,32){.}
\put(276,32){.}
\put(289,288){.}
\put(289,32){.}
\put(289,32){.}
\put(303,290){.}
\put(303,32){.}
\put(303,32){.}
\put(317,291){.}
\put(317,32){.}
\put(317,32){.}
\put(5,205){0.3}
\put(325,15){5}
\put(25,25){\vector(0,1){300}}
\put(25,25){\vector(1,0){300}}
\put(15,340){$r_1, r_2, r_3$}
\put(330,30){$l$}
\end{picture}

\begin{center}
Fig.3
\end{center}

{ \it
 The relative errors of $N_A$, $N_B$ and $N_C$
drawn as the function of $l$ at $b=0$.
Parameter $l$ goes from $0.01$ up to $5.01$.

One can see
two wings which
come together for $b$ near $0$. The upper wing corresponds to
the relative error of $N_A$. The lower wing corresponds to the relative
errors of $N_B$ and $N_C$. One can not separate them.

One can see that $r_1$ decreases when $b$ goes to $0$ and can estimate
$r_2$ by $r_1$ at small $l$.

}

\pagebreak

\begin{picture}(350,350)
\put(0,0){\line(0,1){350}}
\put(0,0){\line(1,0){350}}
\put(350,0){\line(0,1){350}}
\put(0,350){\line(1,0){350}}
\put(26,191){.}
\put(26,177){.}
\put(27,175){.}
\put(27,175){.}
\put(28,175){.}
\put(29,175){.}
\put(29,175){.}
\put(30,175){.}
\put(30,175){.}
\put(31,175){.}
\put(32,175){.}
\put(33,175){.}
\put(35,175){.}
\put(36,175){.}
\put(37,175){.}
\put(38,175){.}
\put(39,175){.}
\put(41,175){.}
\put(42,175){.}
\put(43,175){.}
\put(45,175){.}
\put(47,175){.}
\put(48,175){.}
\put(50,175){.}
\put(52,175){.}
\put(54,175){.}
\put(56,175){.}
\put(58,175){.}
\put(60,175){.}
\put(63,175){.}
\put(65,175){.}
\put(68,175){.}
\put(71,175){.}
\put(74,175){.}
\put(77,175){.}
\put(80,175){.}
\put(83,175){.}
\put(87,175){.}
\put(90,175){.}
\put(94,175){.}
\put(98,175){.}
\put(102,175){.}
\put(107,175){.}
\put(111,175){.}
\put(116,175){.}
\put(122,175){.}
\put(127,175){.}
\put(133,175){.}
\put(139,175){.}
\put(146,175){.}
\put(152,175){.}
\put(159,175){.}
\put(167,175){.}
\put(174,175){.}
\put(182,175){.}
\put(191,175){.}
\put(200,175){.}
\put(209,175){.}
\put(219,175){.}
\put(229,175){.}
\put(240,175){.}
\put(251,175){.}
\put(263,175){.}
\put(276,175){.}
\put(289,175){.}
\put(303,175){.}
\put(317,175){.}
\put(5,175){1}
\put(325,15){5}
\put(25,25){\vector(0,1){300}}
\put(25,25){\vector(1,0){300}}
\put(15,340){$r_2 / r_3$}
\put(330,30){$l$}
\end{picture}

\begin{center}
Fig.4
\end{center}

{ \it
 The ratio $r_2 / r_3$
drawn as the function of $l$ at $b=0$.
Parameter $l$ goes from $0.01$ up to $5.01$.

One can see the difference between $r_2$ and $r_3$ only at very
small values of $b$ (only the first two points corresponding to $l=0.01$
and $l=0.02$). So, $N_3$ can be also considered as a suitable approximation.

}

\end{document}